\documentstyle[twoside,psfig]{article}

\catcode`\@=11
\long\def\@makefntext#1{
\protect\noindent \hbox to 3.2pt {\hskip-.9pt  
$^{{\eightrm\@thefnmark}}$\hfil}#1\hfill}		

\def\@makefnmark{\hbox to 0pt{$^{\@thefnmark}$\hss}}	
	
\def\ps@myheadings{\let\@mkboth\@gobbletwo
\def\@oddhead{\hbox{}
\rightmark\hfil\eightrm\thepage}   
\def\@oddfoot{}\def\@evenhead{\eightrm\thepage\hfil
\leftmark\hbox{}}\def\@evenfoot{}
\def\sectionmark##1{}\def\subsectionmark##1{}}

\oddsidemargin=\evensidemargin
\newcounter{sectionc}\newcounter{subsectionc}\newcounter{subsubsectionc}
\renewcommand{\section}[1] {\vspace{12pt}\addtocounter{sectionc}{1} 
\setcounter{subsectionc}{0}\setcounter{subsubsectionc}{0}\noindent 
	{\tenbf\thesectionc. #1}\par\vspace{5pt}}
\renewcommand{\subsection}[1] {\vspace{12pt}\addtocounter{subsectionc}{1} 
	\setcounter{subsubsectionc}{0}\noindent 
	{\bf\thesectionc.\thesubsectionc. {\kern1pt \bfit #1}}\par\vspace{5pt}}
\renewcommand{\subsubsection}[1] {\vspace{12pt}\addtocounter{subsubsectionc}{1}
	\noindent{\tenrm\thesectionc.\thesubsectionc.\thesubsubsectionc.
	{\kern1pt \tenit #1}}\par\vspace{5pt}}
\newcommand{\nonumsection}[1] {\vspace{12pt}\noindent{\tenbf #1}
	\par\vspace{5pt}}

\newcounter{appendixc}
\newcounter{subappendixc}[appendixc]
\newcounter{subsubappendixc}[subappendixc]
\renewcommand{\thesubappendixc}{\Alph{appendixc}.\arabic{subappendixc}}
\renewcommand{\thesubsubappendixc}
	{\Alph{appendixc}.\arabic{subappendixc}.\arabic{subsubappendixc}}

\renewcommand{\appendix}[1] {\vspace{12pt}
        \refstepcounter{appendixc}
        \setcounter{figure}{0}
        \setcounter{table}{0}
        \setcounter{lemma}{0}
        \setcounter{theorem}{0}
        \setcounter{corollary}{0}
        \setcounter{definition}{0}
        \setcounter{equation}{0}
        \renewcommand{\thefigure}{\Alph{appendixc}.\arabic{figure}}
        \renewcommand{\thetable}{\Alph{appendixc}.\arabic{table}}
        \renewcommand{\theappendixc}{\Alph{appendixc}}
        \renewcommand{\thelemma}{\Alph{appendixc}.\arabic{lemma}}
        \renewcommand{\thetheorem}{\Alph{appendixc}.\arabic{theorem}}
        \renewcommand{\thedefinition}{\Alph{appendixc}.\arabic{definition}}
        \renewcommand{\thecorollary}{\Alph{appendixc}.\arabic{corollary}}
        \renewcommand{\theequation}{\Alph{appendixc}.\arabic{equation}}
        \noindent{\tenbf Appendix \theappendixc #1}\par\vspace{5pt}}
\newcommand{\subappendix}[1] {\vspace{12pt}
        \refstepcounter{subappendixc}
        \noindent{\bf Appendix \thesubappendixc. {\kern1pt \bfit #1}}
	\par\vspace{5pt}}
\newcommand{\subsubappendix}[1] {\vspace{12pt}
        \refstepcounter{subsubappendixc}
        \noindent{\rm Appendix \thesubsubappendixc. {\kern1pt \tenit #1}}
	\par\vspace{5pt}}

\newcommand{\textlineskip}{\baselineskip=13pt}
\newcommand{\smalllineskip}{\baselineskip=10pt}

\def\eightcirc{
\begin{picture}(0,0)
\put(4.4,1.8){\circle{6.5}}
\end{picture}}
\def\eightcopyright{\eightcirc\kern2.7pt\hbox{\eightrm c}} 

\newcommand{\copyrightheading}[1]
	{\vspace*{-2.5cm}\smalllineskip{\flushleft
	{\footnotesize Modern Physics Letters A #1}\\
	{\footnotesize $\eightcopyright$\, World Scientific Publishing
	 Company}\\
	 }}

\newcommand{\publisher}[2]{{\begin{center}\footnotesize\smalllineskip 
	Received #1\\
	Revised #2
	\end{center}
	}}

\def\abstracts#1#2#3{{
	\centering{\begin{minipage}{4.5in}\footnotesize\baselineskip=10pt
	\parindent=0pt #1\par 
	\parindent=15pt #2\par
	\parindent=15pt #3
	\end{minipage}}\par}} 

\def\keywords#1{{
	\centering{\begin{minipage}{4.5in}\footnotesize\baselineskip=10pt
	{\footnotesize\it Keywords}\/: #1
	 \end{minipage}}\par}}

\newcommand{\bibit}{\nineit}
\newcommand{\bibbf}{\ninebf}
\renewenvironment{thebibliography}[1]
	{\frenchspacing
	 \ninerm\baselineskip=11pt
	 \begin{list}{\arabic{enumi}.}
        {\usecounter{enumi}\setlength{\parsep}{0pt}     
	 \setlength{\leftmargin 12.7pt}{\rightmargin 0pt} 
         \setlength{\itemsep}{0pt} \settowidth
	{\labelwidth}{#1.}\sloppy}}{\end{list}}

\newcounter{itemlistc}
\newcounter{romanlistc}
\newcounter{alphlistc}
\newcounter{arabiclistc}

\newcommand{\fcaption}[1]{
        \refstepcounter{figure}
        \setbox\@tempboxa = \hbox{\footnotesize Fig.~\thefigure. #1}
        \ifdim \wd\@tempboxa > 5in
           {\begin{center}
        \parbox{5in}{\footnotesize\smalllineskip Fig.~\thefigure. #1}
            \end{center}}
        \else
             {\begin{center}
             {\footnotesize Fig.~\thefigure. #1}
              \end{center}}
        \fi}

\newcommand{\tcaption}[1]{
        \refstepcounter{table}
        \setbox\@tempboxa = \hbox{\footnotesize Table~\thetable. #1}
        \ifdim \wd\@tempboxa > 5in
           {\begin{center}
        \parbox{5in}{\footnotesize\smalllineskip Table~\thetable. #1}
            \end{center}}
        \else
             {\begin{center}
             {\footnotesize Table~\thetable. #1}
              \end{center}}
        \fi}

\def\@citex[#1]#2{\if@filesw\immediate\write\@auxout
	{\string\citation{#2}}\fi
\def\@citea{}\@cite{\@for\@citeb:=#2\do
	{\@citea\def\@citea{,}\@ifundefined
	{b@\@citeb}{{\bf ?}\@warning
	{Citation `\@citeb' on page \thepage \space undefined}}
	{\csname b@\@citeb\endcsname}}}{#1}}

\newif\if@cghi
\def\cite{\@cghitrue\@ifnextchar [{\@tempswatrue
	\@citex}{\@tempswafalse\@citex[]}}
\def\citelow{\@cghifalse\@ifnextchar [{\@tempswatrue
	\@citex}{\@tempswafalse\@citex[]}}
\def\@cite#1#2{{$\null^{#1}$\if@tempswa\typeout
	{IJCGA warning: optional citation argument 
	ignored: `#2'} \fi}}

\def\pmb#1{\setbox0=\hbox{#1}
	\kern-.025em\copy0\kern-\wd0
	\kern.05em\copy0\kern-\wd0
	\kern-.025em\raise.0433em\box0}

\def\fnm#1{$^{\mbox{\scriptsize #1}}$}
\def\fnt#1#2{\footnotetext{\kern-.3em
	{$^{\mbox{\scriptsize #1}}$}{#2}}}

\def\ps@myheadings{%
    \let\@oddfoot\@empty\let\@evenfoot\@empty
    \def\@evenhead{\slshape\leftmark\hfil}
    \def\@oddhead{\hfil{\slshape\rightmark}}
    \let\@mkboth\@gobbletwo
    \let\sectionmark\@gobble
    \let\subsectionmark\@gobble
    }
\font\tenrm=cmr10
\font\tenit=cmti10 
\font\tenbf=cmbx10
\font\bfit=cmbxti10 at 10pt
\font\ninerm=cmr9
\font\nineit=cmti9
\font\ninebf=cmbx9
\font\eightrm=cmr8

\def\qed{\hbox{${\vcenter{\vbox{			
   \hrule height 0.4pt\hbox{\vrule width 0.4pt height 6pt
   \kern5pt\vrule width 0.4pt}\hrule height 0.4pt}}}$}}

\newcommand{\be}{\begin{equation}}
\newcommand{\ee}{\end{equation}}
\newcommand{\ba}{\begin{array}}
\newcommand{\ea}{\end{array}}

\newcommand{\slashs}[1]{\not{\!#1}}

\def\bea{\begin{eqnarray}}
\def\eea{\end{eqnarray}}

\begin{document}
\setlength{\textheight}{7.7truein}  

\markboth{\protect{\footnotesize\it Transverse Momentum Distribution in the $B$ Mesons
}}{\protect{\footnotesize\it Transverse Momentum Distribution in the $B$ Mesons
}}

\normalsize\textlineskip

\setcounter{page}{1}

\copyrightheading{}	

\vspace*{0.88truein}

\centerline{\bf TRANSVERSE MOMENTUM DISTRIBUTION IN THE $B$ MESONS}
\baselineskip=13pt
\centerline{\bf IN THE HEAVY-QUARK LIMIT:
THE WANDZURA-WILCZEK PART
}
\vspace*{0.4truein}
\centerline{\footnotesize HIROYUKI KAWAMURA
}
\baselineskip=12pt
\centerline{\footnotesize\it Deutsches Elektronen-Synchrotron, DESY}
\baselineskip=10pt
\centerline{\footnotesize\it Platanenallee 6, D 15738 Zeuthen, Germany
}
\vspace*{12pt}

\centerline{\footnotesize JIRO KODAIRA\, and CONG-FENG QIAO\footnote{JSPS Research Fellow.}}
\baselineskip=12pt
\centerline{\footnotesize\it Department of Physics, Hiroshima University}
\baselineskip=10pt
\centerline{\footnotesize\it Higashi-Hiroshima 739-8526, Japan}
\vspace*{12pt}

\centerline{\footnotesize KAZUHIRO TANAKA}
\baselineskip=12pt
\centerline{\footnotesize\it Department of Physics, Juntendo University}
\baselineskip=10pt
\centerline{\footnotesize\it Inba-gun, Chiba 270-1695, Japan}
\vspace*{0.228truein}

\publisher{(received date)}{(revised date)}

\vspace*{0.23truein}
\abstracts{In the heavy-quark limit,
the valence
Fock-state components in the $B$ mesons
are described by a set of two light-cone wavefunctions.
We show that these two wavefunctions
obey simple
coupled differential equations, which are
based on the equations of motion
in the Heavy Quark Effective Theory (HQET),
and the analytic solutions for them are obtained.
The results generalize
the recently obtained longitudinal-momentum distribution
in the Wandzura-Wilczek
approximation
by including the transverse momenta.
We find that the transverse momentum distribution
depends on the longitudinal momentum of the constituents, and that the
wavefunctions
damp very slowly for large transverse separation between quark and
antiquark.}{}{}

\vspace*{10pt}

\vspace*{2pt}

\textlineskip			
\vspace*{12pt}			

\baselineskip=13pt	        
\normalsize              	
\noindent
Along with the progress in both theory and experiment, $B$ physics
becomes one of the most active research areas in high energy
physics.
Many $B$ meson
exclusive decay processes 
turn out to be calculable 
systematically in the frameworks of
newly developed factorization formalisms,
the so-called pQCD
approach \cite{h.n.li,kls} and QCD Factorization
approach.\cite{Beneke:2000ry,Beneke:2001ev,Bauer:2001cu}
In all of the calculations based on the factorization approaches,
the light-cone distribution amplitudes of the participating mesons,
which 
express the nonperturbative long-distance contributions 
in the factorized amplitudes,\cite{Chernyak:1984ej,bl} play an important role
in making reliable predictions.
The light-cone distribution amplitudes
describe the probability amplitudes 
to find particular partons with definite light-cone momentum
fraction in a meson, and thus are process-independent quantity.
It is well-known that, for the light mesons, the
model-independent framework to construct the light-cone distribution
amplitudes is established
for 
leading and higher twists as well.\cite{Braun:1990iv,Ball:1998sk}
However, unfortunately, the distribution amplitudes for
the $B$ mesons are 
not well-known at present and they provide a major
source of uncertainty in the calculations of the decay rates.

Recently,\cite{kkqt} we have presented
the first systematic study for the $B$ meson light-cone distribution
amplitudes,
and 
derived explicit forms for the quark-antiquark distribution
amplitudes,
which exactly satisfy
the constraints coming from the equations of motion and heavy-quark symmetry.
We have found that the ``Wandzura-Wilczek-type'' contributions 
which correspond to the valence quark distributions,
are determined uniquely in analytic form in terms of $\bar{\Lambda}$,
a fundamental mass parameter of Heavy Quark Effective 
Theory (HQET).\cite{Isgur:1989vq,Neubert:1994mb}
We have also shown that both leading- and higher-twist distributions
receive the contributions 
from the multi-particle states with additional dynamical gluons,
and derived the exact integral representations
for these contributions.

By definition, the light-cone distribution amplitudes
are given by the light-cone (Bethe-Salpeter) wavefunctions
at zero transverse separation of the constituents.
Thus the previous results of Ref.\cite{kkqt} have been obtained for
the configuration in which the quark and antiquark are separated
by exactly light-like distance.
The information on transverse momentum distribution has been
integrated out. 
However, the light-cone wavefunctions with transverse momentum dependence
are necessary for computing the power corrections to the exclusive
amplitudes,
and also
for estimating the transition form factors for $B \rightarrow D$, $B
\rightarrow \pi$, etc,
which constitute another type of long-distance contributions appearing in
the factorization
approaches for the exclusive $B$ meson decays.

In this Letter, we extend the analysis of Ref.\cite{kkqt}
to include the
transverse momentum effects. In particular, we 
derive explicit analytic
formulae for a complete set of the $B$ meson light-cone wavefunctions
within 
the Wandzura-Wilczek approximation.\cite{kkqt}
It should be noted that the Wandzura-Wilczek approximation
is not equivalent to the free field approximation.
The leading Fock-states, which correspond to the twist-2 
contribution in the case of light meson wavefunctions,
carries the effect from the QCD interaction.\cite{Braun:1990iv,Ball:1998sk}
We also 
estimate the effects neglected in this ``valence'' approximation;
we derive the exact result for the first moment of 
the transverse momentum squared $\mbox{\boldmath $k$}_{T}^{2}$
in terms of the full light-cone wavefunctions,
which include the higher Fock-states
with additional dynamical, nonperturbative gluons.\fnm{a}
\fnt{a}{Radiative corrections due to the virtual gluons and/or quark-antiquark
pairs can be included as the renormalization scale-depenence of the wavefunctions,
which is governed by the renormalization group equations for the nonlocal operators 
in Eq. (\ref{phi}) below.
The discussion of this point is beyond the scope of this work.}

The light-cone wavefunctions are related to the usual Bethe-Salpeter
wavefunctions at equal
light-cone time $z^{+}= (z^{0}+z^{3})/\sqrt{2}$.\cite{bl}
The quark-antiquark light-cone wavefunctions $\tilde{\psi}_{\pm}(t,z^2)$ 
of the $B$ mesons
in the heavy-quark limit are defined by the
vacuum-to-meson matrix element of nonlocal
operators, 
following Refs.\cite{sachrajda,Pirjol:2000gn}:
\bea
\lefteqn{\langle 0 | \bar{q}(z) \Gamma h_{v}(0) |\bar{B}(p) \rangle}\nonumber\\
 &=& - \frac{i f_{B} M}{2} {\rm Tr}
 \left[ \gamma_{5}\Gamma \frac{1 + \slashs{v}}{2}
 \left\{ \tilde{\psi}_{+}(t,z^2) - \slashs{z} \frac{\tilde{\psi}_{+}(t,z^2)
 -\tilde{\psi}_{-}(t,z^2)}{2t}\right\} \right]\ .
 \label{phi}
\eea
Here $z^{+}=0$, $v^{2} = 1$, $t=v\cdot z$, and $p^{\mu} = Mv^{\mu}$
is the 4-momentum of the $B$ meson with mass $M$.
$h_{v}(x)$ denotes the effective $b$-quark field,
$b(x) \approx \exp(-im_{b} v\cdot x)h_{v}(x)$,
and is subject to the on-shell constraint,
$\slashs{v} h_{v} = h_{v}$.\cite{Isgur:1989vq,Neubert:1994mb}
$\Gamma$ is a generic Dirac matrix and,
here and in the following, the path-ordered gauge
factors are implied in between the constituent fields.
$f_{B}$ is the decay constant defined as usually as
\bea
\langle 0 | \bar{q}(0) \gamma^{\mu}\gamma_{5} h_{v}(0) |\bar{B}(p) \rangle
   = i f_{B} M v^{\mu}\ ,
\label{fb}
\eea
so that $\tilde{\psi}_{\pm}(t=0,z^2=0) = 1$.
Eq. (\ref{phi}) is the most general parameterization compatible with Lorentz
invariance and the heavy-quark limit.

Note that in the definition (\ref{phi})
the separation $z^{\mu}=(0, z^{-}, \mbox{\boldmath $z$}_{T})$
between quark and antiquark is not restricted
on the light-cone, $z^{2}= - \mbox{\boldmath $z$}_{T}^{2}$,
unlike the corresponding definition
of the distribution amplitudes.\cite{kkqt,Grozin:1997pq,Beneke:2001wa}
Thus the distribution amplitudes $\tilde{\phi}_{\pm}(t)$ in the notation
of Refs.\cite{kkqt,Grozin:1997pq,Beneke:2001wa} are
given by the light-cone limit of the above wavefunctions as
$\tilde{\phi}_{\pm}(t)=\left. \tilde{\psi}_{\pm}(t, z^{2})\right|_{z^{2}
\rightarrow 0}$.

We also introduce the Fourier transforms
with respect to the longitudinal separation $t$ by
\begin{equation}
 \tilde{\psi}_{\pm}(t, z^{2}) = \int d\omega \ e^{-i \omega t}
  \psi_{\pm}(\omega, z^{2}) \ . \label{mom}
\end{equation}
Here $\omega v^{+}$ has the meaning of the light-cone projection
$k^{+}$ of the light-antiquark momentum in the $B$ meson.

We now exploit the constrains from the equations of motion.
The 
procedure here is completely in parallel with that of our previous work,
so we refer the readers to Ref.\cite{kkqt} for the detail.
The matrix elements of the exact operator identities
(Eqs. (3), (4) of Ref.\cite{kkqt}) from the equations
of motion yield a system of four differential equations:
\bea
\label{eq:1}
\omega \frac{\partial \psi_{-}}{\partial \omega}
&+& z^2 \left(\frac{\partial \psi_{+}}{\partial z^2}
-\frac{\partial \psi_{-}}{\partial z^2}\right)
+ \psi_{+}
= 0\;, \\
\label{eq:2}
\omega \left(\frac{\partial \psi_{+}}{\partial \omega} -
\frac{\partial \psi_{-}}{\partial \omega}\right)
&+& 4  \frac{\partial^{2}}{\partial \omega^{2}}\frac{\partial
\psi_{+}}{\partial z^2}
+2 \left(\psi_{+}-\psi_{-}\right)
= 0\;, \\
\label{eq:3}
(\omega - \bar{\Lambda}) \frac{\partial \psi_{+}}{\partial \omega} &+&
2  \frac{\partial^{2}}{\partial \omega^{2}}\frac{\partial \psi_{+}}{\partial
z^2}
+\frac{1}{2}\left(3\psi_{+}-\psi_{-}\right)
= 0\;,\\
\label{eq:4}
(\omega - \bar{\Lambda}) \left(\frac{\partial \psi_{+}}{\partial \omega}
- \frac{\partial \psi_{-}}{\partial \omega}\right)
 &+&
2  \frac{\partial^{2}}{\partial \omega^{2}}
\left(\frac{\partial \psi_{+}}{\partial z^2} -
\frac{\partial \psi_{-}}{\partial z^2} \right)
+ 2 \left(\psi_{+} - \psi_{-}\right)
= 0\;.
\eea
Here,
\begin{equation}
  \bar{\Lambda} = M - m_{b} =
 \frac{iv\cdot \partial \langle 0| \bar{q} \Gamma h_{v} |\bar{B}(p) \rangle}
  {\langle 0| \bar{q} \Gamma h_{v} |\bar{B}(p) \rangle}
\label{lambda}
\end{equation}
is the usual ``effective mass'' of meson states in the HQET,\cite{Neubert:1994mb,Falk:1992fm} 
and the shorthand notation
$\psi_{\pm} \equiv \psi_{\pm}(\omega, z^2)$ is adopted.
These Eqs. (\ref{eq:1}), (\ref{eq:2}), (\ref{eq:3}), and (\ref{eq:4})
correspond to Eqs. (7), (8), (10), and (11) in Ref.\cite{kkqt},
respectively,
but are given in the ``$\omega$-representation'' instead of the
``$t$-representation''
via Eq. (\ref{mom}).
Further differences compared with Ref.\cite{kkqt} are
on two aspects.
Firstly, the light-cone limit is not taken
in order to explore the tranverse momentum
distribution, so that the terms proportional
to $z^{2}$  appear in the above Eq. (\ref{eq:1}).
Secondly, we restrict
our interest within the two-particle Fock states
neglecting the contribution 
from the quark-antiquark-gluon
three-particle operators.

{}From Eqs. (\ref{eq:2}) and (\ref{eq:3}), we obtain
\begin{equation}
\left(\omega - 2 \bar{\Lambda} \right)
\frac{\partial \psi_{+}}{\partial \omega} + \omega \frac{\partial
\psi_{-}}{\partial \omega}
+ \psi_{+} + \psi_{-} = 0 \; .
\label{eq:t1}
\end{equation}
This can be integrated with boundary conditions
$\psi_{\pm}(\omega, z^{2}) = 0$ for  $\omega <0$ or $\omega \rightarrow
\infty$ as
\begin{equation}
\left( \omega - 2 \bar{\Lambda} \right) \psi_{+} + \omega \psi_{-} = 0\; .
\label{eq:t2}
\end{equation}
A system of Eqs. (\ref{eq:1}) and (\ref{eq:t2}) has been solved for
$z^{2}\rightarrow 0$
and completely determined the wavefunctions in the
light-cone limit.\cite{kkqt} The results are
\bea
 \psi_{+}(\omega, z^{2}=0) &\equiv& \phi_{+}(\omega) = \frac{\omega}{2
\bar{\Lambda}^{2}}
 \theta(\omega) \theta(2 \bar{\Lambda} - \omega) \ , \label{solp} \\
 \psi_{-}(\omega, z^{2}=0) &\equiv& \phi_{-}(\omega) =
  \frac{2 \bar{\Lambda} - \omega}{2 \bar{\Lambda}^{2}}
  \theta(\omega) \theta(2 \bar{\Lambda} - \omega) \ . \label{solm}
\eea
These solutions for $z^{2}=0$
serve as ``boundary conditions''
to solve Eqs. (\ref{eq:1})-(\ref{eq:4}) for $z^{2} \neq 0$.
{}From Eqs. (\ref{eq:1}) and (\ref{eq:t2}) for $z^{2} \neq 0$, we find 
\begin{equation}
\psi_{\pm}(\omega, z^{2}) = \phi_{\pm}(\omega)
\xi \left(z^{2}\omega(2\bar{\Lambda}-\omega)\right)\ ,
\label{eq:chi}
\end{equation}
where $\xi(x)$ is some function of a single variable $x$, and satisfies
$\xi(0) = 1$ due to Eqs. (\ref{solp}), (\ref{solm}).
The functional form of $\xi(x)$ can be easily determined from a remaining
differential equation, e.g., (\ref{eq:3}), which was useless
in the light-cone limit\cite{kkqt} (Eq. (\ref{eq:4}) gives the
result identical to Eq. (\ref{eq:3})). We obtain
$\xi(x) = J_{0}\left(\sqrt{-x}\right)$,
where $J_{0}$ is a (regular) Bessel function, so that
the analytic solution
for the coupled differential equations (\ref{eq:1})-(\ref{eq:4}) is given by
\begin{equation}
\psi_{\pm}(\omega, -\mbox{\boldmath $z$}_{T}^2)
= \phi_{\pm}(\omega)J_{0}\left(|\mbox{\boldmath
$z$}_{T}|\sqrt{\omega(2\bar{\Lambda}-\omega)} \right) \ .
\label{j0}
\end{equation}
These are the light-cone wavefunctions for the transverse separation
$\mbox{\boldmath $z$}_{T}$
between quark and antiquark.

{}For the momentum-space wavefunctions $\psi_{\pm}(\omega, \mbox{\boldmath
$k$}_{T})$
defined by
\begin{equation}
\tilde{\psi}_{\pm}(t, -\mbox{\boldmath $z$}_{T}^2) =
\int d\omega d^{2}k_{T}\
e^{-i\omega t + i\mbox{\boldmath $k$}_{T}\cdot\mbox{\boldmath $z$}_{T}}
\psi_{\pm}(\omega, \mbox{\boldmath $k$}_{T}) \ ,
\label{momw}
\end{equation}
our solution gives
\bea
\label{eq:12}
\psi_{+}(\omega, \mbox{\boldmath $k$}_{T}) &=&
\frac{\omega}
{2 \pi \bar{\Lambda}^2} \theta(\omega)\theta(2 \bar{\Lambda} - \omega)
\delta \left(\mbox{\boldmath $k$}_{T}^{2} - \omega (2 \bar{\Lambda} -
\omega) \right) \; ,\\
\label{eq:13}
\psi_{-}(\omega, \mbox{\boldmath $k$}_{T}) &=&
\frac{2 \bar{\Lambda} - \omega}{2 \pi \bar{\Lambda}^2}
\theta(\omega)\theta(2 \bar{\Lambda} - \omega)
\delta \left(\mbox{\boldmath $k$}_{T}^{2} - \omega (2 \bar{\Lambda} -
\omega) \right) \; .
\eea
The results (\ref{eq:12}) and (\ref{eq:13}) give exact description
of the valence Fock components of the $B$ meson
wavefunctions in the heavy-quark limit, and represent their transverse
momentum dependence explicitly.
These results show that the dynamics within the two-particle Fock states
is determined solely in terms of a single nonperturbative
parameter $\bar{\Lambda}$.

Up to now, the transverse momentum distributions in the $B$ mesons
have been completely unknown,
so that various models have been used in the literature.
Sometimes the dependence of the light-cone wavefunctions on the transverse
separation is simply neglected, such as
$\psi(\omega, -\mbox{\boldmath $z$}_{T}^2)=\psi(\omega, 0)$; clearly,
this in general contradicts our results (\ref{j0}).
Another frequently used models assume
complete separation (factorization)
between the longitudinal and transverse momentum-dependence in the
wavefunctions,
such as $\psi(\omega, \mbox{\boldmath $k$}_{T}) =
\phi(\omega)\tau(\mbox{\boldmath $k$}_{T})$
(see e.g. Refs.\cite{sachrajda,kls,Bauer:fx}).
One typical example of such models is given by\cite{kls}
\begin{equation}
\psi^{KLS}(\omega, \mbox{\boldmath $k$}_{T})
= N\omega^{2}(1-\omega)^{2}\exp\left( -
\frac{\omega^{2}}{2\omega_{0}^{2}}\right)
\times \exp\left(- \frac{\mbox{\boldmath $k$}_{T}^{2}}{2K^{2}}\right)\ ,
\label{eq:model}
\end{equation}
where $N$ is the normalization constant and $\omega_{0} = 0.3$GeV, $K =
0.4$GeV.\fnm{b}
\fnt{b}{In this model, the two independent wavefunctions $\psi_{\pm}$ of
the $B$ mesons
are set equal to each other, $\psi^{KLS}\equiv
\psi_{+}^{KLS}=\psi_{-}^{KLS}$.}
Our results (\ref{eq:12}) and (\ref{eq:13}) show that the dependence on
transverse and longitudinal momenta is strongly 
correlated through the combination
$\mbox{\boldmath $k$}_{T}^{2}/[\omega(2\bar{\Lambda}-\omega)]$,
therefore the ``factorization models''
are not justified.\fnm{c}
\fnt{c}{The ``non-factorization'' in the light-cone 
wavefunctions for the light-mesons 
has been discussed in Ref.\cite{Zhitnitsky:1993vb}, 
where the coupling between  
transverse and longitudinal momenta through the variable 
$\mbox{\boldmath $k$}_{T}^{2}/[u(1-u)]$,
with $u$ the momentum fraction of the light quark, has been demonstrated.}
We further note that many models assume\cite{sachrajda,kls,Bauer:fx}
Gaussian distribution
for the $\mbox{\boldmath $k$}_{T}$-dependence as in Eq. (\ref{eq:model}).
These models 
show strong dumping at 
large $|\mbox{\boldmath $z$}_{T}|$ as $\sim \exp\left(-
K^2\mbox{\boldmath $z$}_{T}^{2}/2\right)$.
In contrast to this asymptotic behavior,
our wavefunctions (\ref{j0}) have slow-damping with oscillatory behavior
as
$\psi_{\pm}(\omega, -\mbox{\boldmath $z$}_{T}^{2})
\sim \cos(|\mbox{\boldmath $z$}_{T}|\sqrt{\omega(2\bar{\Lambda}-\omega)}
-\pi/4)
/\sqrt{|\mbox{\boldmath $z$}_{T}|}$.

To summarize, in this work we have derived a system of differential equations
for the $B$ meson light-cone wavefunctions including the transverse
degrees of freedom, and obtained the exact solution within the valence Fock
states. The differential equations are derived from the exact equations 
of motion of QCD in the heavy-quark limit.
The heavy-quark symmetry plays an essential role 
as in the case of the light-cone limit.\cite{kkqt,Grozin:1997pq,Beneke:2001wa}
Heavy-quark spin symmetry reduces the number of 
independent wavefunctions drastically,  
so that the valence Fock-state components in the $B$ mesons are described by
only two light-cone wavefunctions.
As a result, a system of four differential equations from the 
equations of motion
becomes a ``complete set'' to determine these two wavefunctions, and
this enables us to obtain the exact solution with  
full account of the $\mbox{\boldmath $k$}_{T}$-dependence.
Also due to
the power of heavy-quark symmetry,
our final results are given in simple analytic formulae involving
one single nonperturbative parameter $\bar{\Lambda}$.
Heavy-quark symmetry also guarantees that the solution in the present paper
provides complete description of the light-cone valence Fock wavefunctions
for the $B^{*}$ mesons and also for the $D$, $D^{*}$ mesons in the
heavy-quark limit.

{}Finally a comment is in order concerning the error induced
by the Wandzura-Wilczek approximation.
From the study of the $B$ meson distribution amplitudes in the light-cone
limit,
there has been indication that, in the heavy-light quark systems,
the higher Fock states could play important roles even in the
leading twist level.\cite{kkqt}
This would suggest that the shape of the wavefunctions
as function of momenta and
their quantitative role in the phenomenological applications would be
modified
when including the higher Fock states. 
{}For example, inspecting the $t \rightarrow 0$ limit 
of Eqs. (8) and (11) of Ref.\cite{kkqt},
one immediately obtains the exact result for the first moment of $\mbox{\boldmath $k$}_{T}^{2}$
as 
\begin{equation}
\int d\omega d^{2}k_{T}\ \mbox{\boldmath $k$}_{T}^{2}
\psi_{\pm}^{(tot)}(\omega, \mbox{\boldmath $k$}_{T})
= 4 \left. \frac{\partial \tilde{\psi}_{\pm}^{(tot)}(t=0, z^{2})}{\partial z^{2}}\right|_{z^{2}\rightarrow 0}
= \frac{2}{3}\left(\bar{\Lambda}^{2}+\lambda_{E}^{2}+\lambda_{H}^{2}\right)\ ,
\label{eq:3part}
\end{equation}
where $\psi_{\pm}^{(tot)} \equiv \psi_{\pm}+\psi_{\pm}^{(hF)}$ denote
the total wavefunctions including the higher Fock contributions $\psi_{\pm}^{(hF)}$,
and $\lambda_{E}$ and $\lambda_{H}$ denote the 
reduced matrix elements of relevant quark-antiquark-gluon operators
in the notation of Ref.\cite{kkqt},
representing the chromoelectric and chromomagnetic fields
in the $B$ meson rest frame, respectively.
The first term $\frac{2}{3}\bar{\Lambda}^{2}$ in the RHS 
of Eq. (\ref{eq:3part}) 
coincides with the moment
of our solution $\psi_{\pm}$, Eqs. (\ref{eq:12}) and (\ref{eq:13}), 
in the Wandzura-Wilczek approximation, while 
other terms $\frac{2}{3}\left(\lambda_{E}^{2}+ \lambda_{H}^{2}\right)$ 
come from the higher Fock contributions
$\psi_{\pm}^{(hF)}$.
The result (\ref{eq:3part}), combined with an estimate
$\lambda_{E}^{2}/\bar{\Lambda}^{2} = 0.36 \pm 0.20$,
$\lambda_{H}^{2}/\bar{\Lambda}^{2} = 0.60 \pm 0.23$
by QCD sum rules,\cite{Grozin:1997pq} suggests that the higher Fock 
contributions might considerably broaden the transverse momentum distribution.
This point can be
studied in a systematic way, and 
more sophisticated wavefunctions $\psi_{\pm}^{(tot)}$ will 
be discussed in detail
in a separate publication.\cite{kkqt3}
However, qualitative
features revealed in this paper, like non-factorization of
longitudinal and transverse directions, ``slow-damping'' for transverse
directions, etc.,
will be unaltered by the effects of multi-particle states,
and helpful in elucidating QCD factorization theorems.


\nonumsection{Acknowledgments}
\noindent
The work of J.K. was supported in part by the Monbu-kagaku-sho
Grant-in-Aid for Scientific Research No.C-13640289.
The work of C-F.Q. was supported by the Grant-in-Aid of JSPS committee.

\nonumsection{References}
%

\end{document}